\documentclass[preprint,10pt]{article}
\usepackage{amssymb}
\usepackage{graphicx}
\usepackage{graphics}
\usepackage{bm}
\begin{document}
\title{Phase diagram of mechanically stretched DNA: The salt effect}
\author{\small{Amar Singh, Navin Singh}\\ 
\small{Department of Physics, Birla Institute of Technology \& Science}\\ \small{Pilani - 333 031, 
Rajasthan, India}}
\date{}

\maketitle

\begin{abstract}
The cations, in form of salt, present in the solution containing DNA play a crucial
role in the opening of two strands of DNA. We use a simple non linear model and
investigate the role of these cations on the mechanical unzipping of DNA. 
The Hamiltonian is modified to incoporate the solvent effect and the cations present 
in the solution. We calculate the melting temperature as well as the critical
force that is required to unzip the DNA molecule as a function of salt concentration
of the solution. The phase diagrams are found to be in close agreement with the 
experimental phase diagrams.
\end{abstract}

\section{Introduction}
\label{intro}
The stability of the double stranded DNA (dsDNA) molecule is primarily due to the 
hydrogen bonding present between the bases of the complimentary strand. The bases 
along the strands give rise to the stacking interaction between the nearest base pairs 
which contributes to the rigidness of the molecule. In addition to this, the presence  
of the cations (Na$^{+}$ or Mg$^{2+}$) in the form of salt (in the solution) plays a crucial 
role in the stability of these molecules. The stability of the strand can be monitored by 
changing the temperature, by applying the force on either of the ends or by changing 
the pH of the solution. A systematic investigation done in the past have shown that 
the melting temperature of a dsDNA increases with the salt concentration 
\cite{santa,owc,nik,ritort,hatch,huguet,joyuex2008,krueger}. Since the 
two strands of the dsDNA are negatively charged, to neutralize the Coulombic 
repulsion between the phosphates, the cations like sodium or magnesium ions are 
required. The concentration of these ions contribute not only to the stability of 
the molecule but also play an important role in the folding kinetics of the molecule. 
To understand the mechanism theoretically, the counterion condensation 
model (based on the two state ion distribution) \cite{manning}, the 
Poisson-Boltzmann model (based on mean field calculations) \cite{sharp,grant} 
have been used. Recently, the tight bonding approximation (TBA) \cite{tan}, 
Poland-Scheraga (PS) \cite{ambjoernsson,yeramian} and 
Peyrard Bishop Dauxious (PBD) \cite{nik} models are also used to study the helix coil 
transition in these molecules. Most of these studies focused primarily on the 
thermal stability of the dsDNA molecule as a function of salt concentration of the solution. 

In the recent years, using single molecule force spectroscopy (SMFS) experiments {\it e.g.} 
optical \& magnetic tweezers, atomic force microscope etc. the forces exerted by 
single stranded binding (SSB) proteins in maintaining the open regions of ssDNA has 
been measured directly \cite{ritort,hatch,huguet,joyuex2008}. 
These groups have experimentally measured the force required to destabilize the dsDNA as
a function of concentration of salt in the solution. In addition to this, several groups have also measured 
the presence of the salt on the stretching behavior of DNA \cite{Chaurasiya,hanke,bloomfield,kumar}. 
All these experimental results provide the measurement of these forces as a function of salt concentration.
However, the theoretical understanding of these results is also important in order to get the precise idea 
of the physical processes that are involved in these transition.
In this manuscript, we investigate the effect of salt present in the solution, on the force induced 
unzipping of a heterogeneous dsDNA molecule using PBD model \cite{pb}, which has been discussed in section 2. 
In this section, we also discuss the method to calculate the melting temperature ($T_m$) and the 
forces required to unzip the chain. The method developed in section 2, has been extended to study 
the thermal \& force induced melting of dsDNA in section 3 \& 4, respectively. Section 5 summarizes 
the results followed by brief conclusions.

\section{The model}
In this section, we briefly discuss the basic features of the PBD model, which considers the 
stretching between corresponding bases only. Unlike the PS model, which is based 
on the two state model (bound segment or unbound segment), the PBD model includes 
intermediate state because the stretching is a continuously varying variable.
Although the model ignores the helicoidal structure \cite{saul,cocco,ns4,zoli,pey2009} of 
the dsDNA molecule. It has enough details to analyze mechanical behavior at few ${\rm \AA}$ 
scale relevant to molecular-biological events. The Hamiltonian 
for the considered system of $N$ base pairs unit is written as,
\begin{equation}
\label{eqn1}
H = \sum_{i=1}^N\left[\frac{p_i^2}{2m} + V_S(y_i,y_{i+1}) + V_M(y_i) + V_{\rm sol}(y_i)\right]
\end{equation}
where $y_i$ represents the stretching from the equilibrium position of the hydrogen bonds, 
$p_i = m${{\.y}$_i$} represents the momentum while $m$ is the reduced mass of a base pair 
(taken to be the same for both A-T and  G-C base pairs). The stacking interaction 
between two consecutive base pairs along the chain is represented by, 
\begin{equation}
\label{eqn3}
V_S(y_i,y_{i+1}) = \frac{k}{2}(y_i - y_{i+1})^2[1 + \rho e^{-b(y_i + y_{i+1})}],
\end{equation}
where $k$ represents the single strand elasticity. The anharmonicity in the strand 
elasticity is represented by $\rho$ while $b$ represents its range. These parameters
are assumed to be independent of sequence heterogeneity. The sequence heterogeneity
has effect on the stacking interaction along the strand. This can be taken care 
through the single strand elasticity parameter $k$. 

The hydrogen bonding between the two bases in the $i^{\rm th}$ pair is represented by 
the Morse potential,
\begin{equation}
\label{eqn2}
V_M(y_i) = D_i(e^{-a_iy_i} - 1)^2,
\end{equation}
where $D_i$ represents the potential depth, roughly equal to the bond energy of that pair 
and $a_i$ represents the inverse of the width of the potential well. The
heterogeneity in the sequence is taken care by the values of $D_i$ and $a_i$. 
In the stability of the dsDNA molecule 
the role of hydrogen bond is the key factor. In most of the previous studies, 
the hydrogen bond interaction and the effects of surroundings, such as salt concentration 
of the solution, are taken as constant \cite{pb,saul}. As the DNA molecules are strong 
polyelectrolytes, having negatively charged phosphate groups, it would be interesting to
analyze its role in the melting or unzipping profiles. The salts present in the solution neutralize 
the negative charge of the phosphate groups, therefore, the increase in their concentration will 
reduces the electrostatic repulsive forces between these negatively charged groups. 
Since system at higher concentration prefers to be in less entropic state, more thermal or 
mechanical energy will be required to break the hydrogen bonds. In the PBD model, the stability 
in hydrogen bond is
represented by the depth of Morse potential, $D_i$. Thus, this parameter should be a function of salt
concentration of the solution. Experimental observations predict that 
the melting temperature of dsDNA scales logarithmically with the salt 
present in the solution\cite{owc,blake}. In addition to this, the melting temperature has been 
found to have a 
linear dependence on the value of potential depth. Keeping these factors in the background, 
we modify the potential depth as,
\begin{equation}
\label{eqn4}
D_i = D_0\left[1 + \lambda \ln \left(\frac{C}{C_0}\right)\right]
\end{equation}
Here, the concentration, $C$ is expressed in moles per liter and $C_0$ is the reference 
concentration chosen to be 1 mole/liter. The $\lambda$ appearing in the 
potential is a solution constant \cite{nik,dong}.

An additional term in the Hamiltonian is the solvent term which simulates the formation 
of hydrogen bonds with the solvent, once the hydrogen bonds are stretched by more than 
their equilibrium values. We adopts the solvent term from the refs. \cite{zhang,drukker}. 
\begin{equation}
\label{eqn5}
V_{sol}(y_i) = -\frac{1}{4}D_i\left[\tanh\left(\gamma y_i\right)-1\right]
\end{equation}
The \textquotedblleft$\tanh$\textquotedblright term in the potential enhances the energy 
of the equilibrium configuration 
and the height of the barrier below which the base pair is closed. The small barrier 
basically determines the threshold stretching of hydrogen bond about which a base pair 
may be temporarily broken, re-bonded and then fully broken. Of course, this come to the 
broken state at a length greater than $\sim 2 {\rm \AA}$. As solvent role is to stabilize 
the denatured state, this form of potential can be a good choice. The term, $\gamma$ is the solvent 
interaction factor and it reduces the height of the barrier appears in the potential
\cite{zhang,drukker,weber,zoli2011}. We tune various values of $\gamma$ from 0.1 to 1.0 and plotted the 
effective potential, as shown in figure 1 B. We found that for larger values of $\gamma$, unzipping
transition is more favorable. As the broken state occurs $\sim 2.0 \;{\rm \AA}$, the value of $\gamma$ 
should be chosen which reflect the breaking around $ 2.0 \; {\rm \AA}$. We found  $\gamma = 1.0 \; {\rm \AA^{-1}}$
as a suitable choice for our calculations. 
\begin{figure}[h]
\label{fig01}
\begin{center}
\includegraphics[height=2.5in,width=2.63in]{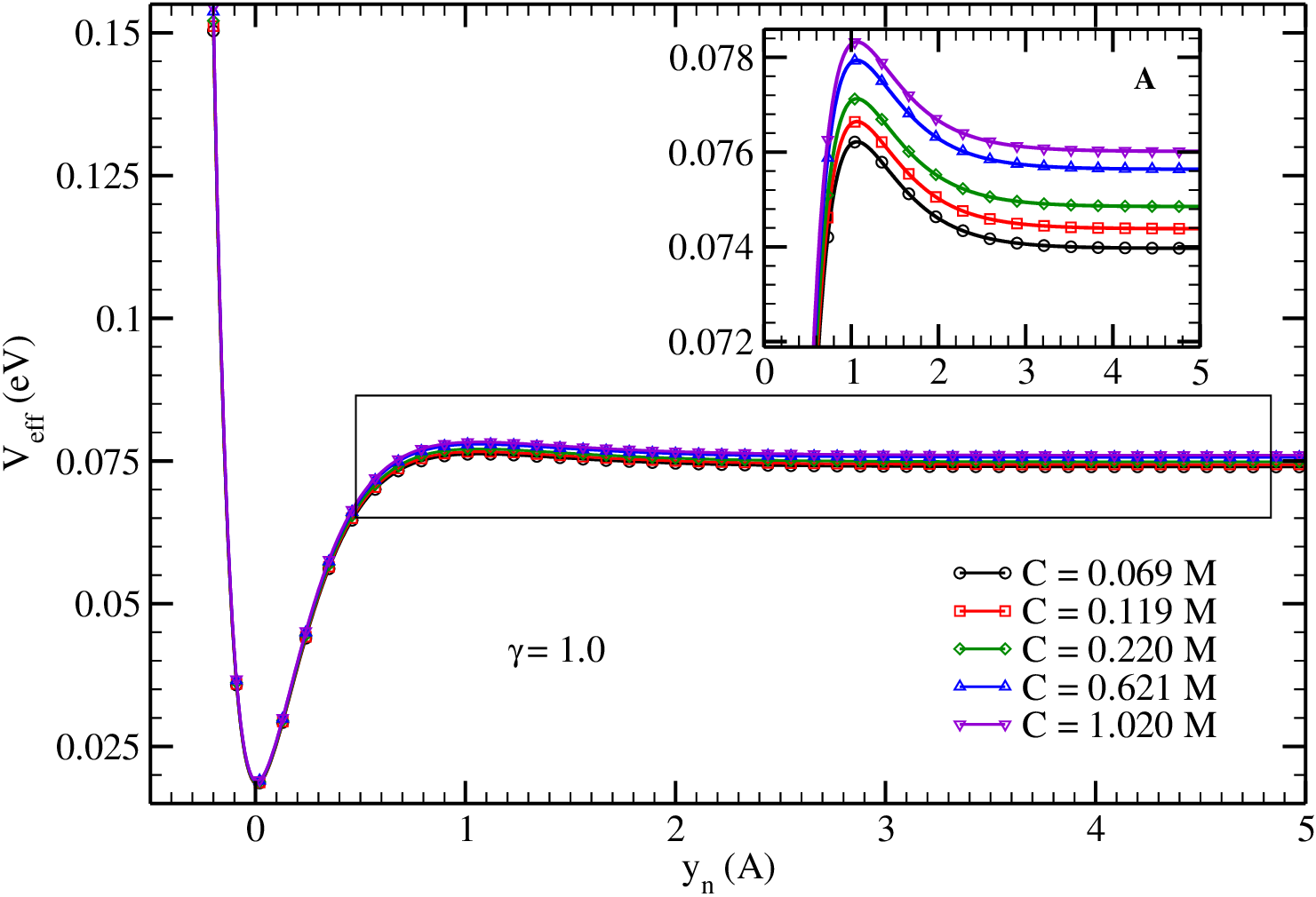}\hspace{0.15cm}
\includegraphics[height=2.5in,width=2.63in]{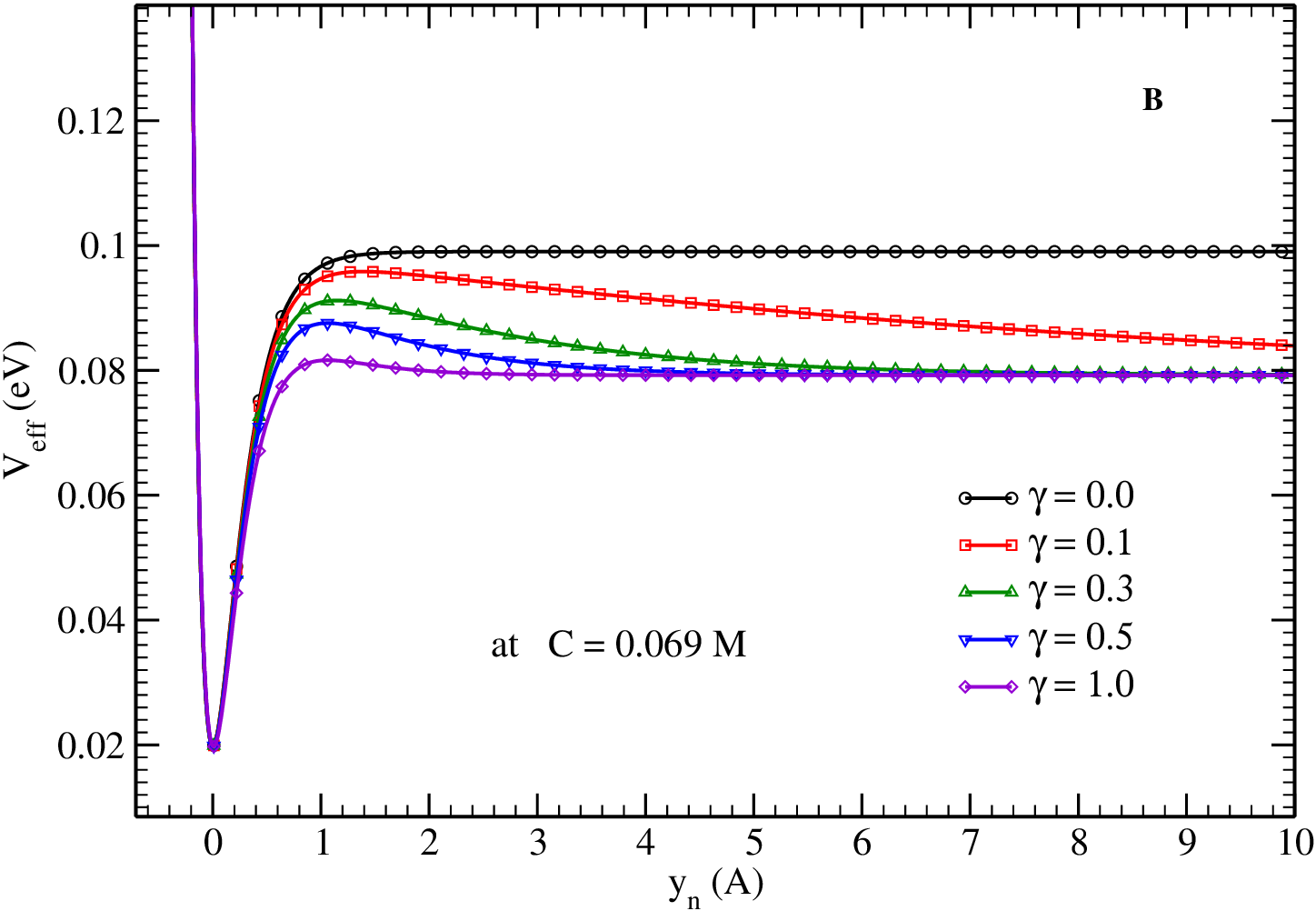}
\caption{Plot of effective potential $V_{\rm eff} = V_M + V_{\rm sol}$ as a function of
base pair stretching (in ${\rm \AA}$). In figure A, the effect of increase in the salt 
concentration of the solution on the potential depth is shown. 
The variation in the barrier height with the increase in the solvent 
interaction factor $\gamma$ is shown in figure B. }
\end{center}
\end{figure}

Thermodynamics of the transition can be investigated by evaluating the expression for the 
partition function. The canonical partition function is written as,
\begin{equation}
\label{eqn6}
Z = \int \prod_{i=1}^{N}\left\{dy_idp_i\exp[-\beta H(y_i,y_{i+1})]\right\} = 
Z_pZ_c,
\end{equation}
where $Z_p$ corresponds to the momentum part of the partition function while the $Z_c$
contributes as the configurational part of the partition function. Since 
the momentum part is decoupled in the integration, it can be integrated out as a 
simple Gaussian integral. This will contribute a factor of $(2\pi mk_BT)^{N/2}$ in the 
partition function, where $N$ is the number of base pairs in the chain. The configurational 
partition function, $Z_c$, is defined as \cite{ns3},
\begin{equation}
\label{eqn7}
Z_c = \int \prod_{i=1}^N dy_i  K(y_i,y_{i+1})
\end{equation}
where $K(y_i,y_{i+1}) = \exp\left[-\beta H(y_i,y_{i+1})\right].$
For the homogeneous chain, one can evaluate the partition function by transfer integral (TI)
method by applying the periodic boundary condition. In case of heterogeneous chain, with open 
boundary, the configurational part of the partition function can be integrated numerically with 
the help of matrix multiplication method. Once the  limit of integration has been chosen, the 
task is reduced to discretized the space to evaluate the integral numerically. We choose 
the limits as $-5.0 \; {\rm \AA}\; {\rm to}\; 200.0 \; {\rm \AA}$, as the lower and upper limits of 
the integration, respectively. The space is being discretized using the Gaussian quadrature formula 
with number of grid points equal to 900. In our previous studies \cite{ns3}, we observed that to get 
precise value of melting temperature ($T_m$) one has to choose the large grid points. We found 
that 900 is quite sufficient number for this purpose. As all matrices in eq.\ref{eqn7} are identical 
in nature the multiplication is done very efficiently. The thermodynamic quantities of interest 
can be calculated by evaluating the Helmholtz free energy of the system. The free energy per 
base pair is,
\begin{equation}
\label{eqn8}
f(T) = -\frac{1}{2}k_B T\ln\left(2\pi m k_B T\right) - \frac{k_B T}{N}\ln Z_c.
\end{equation}
The other thermodynamic quantities like specific heat ($C_v$) is evaluated using the following
relations,
\begin{equation}
\label{eqn9}
C_v(T) = -T\frac{\partial^2 f}{\partial T^2}.
\end{equation}
We also monitor the fraction of open pairs as a function of temperature and force. 
The details of process of separation of short chains are different from that of the 
long dsDNA chains. For long chains, when the fraction $\phi$ of open base pairs goes 
practically from 0 to 1 at the melting transition, the two strands are not yet completely 
separated. At this point, the great majority of the bonds is disrupted and the 
dsDNA has denaturated, but the few bonds still remain intact, prevent the two strands 
going apart from each other. The real separation occurs only at high temperatures.
For very long chains, the double strand is always a single macromolecule, 
and hence one need to calculate the fraction of intact or broken base pairs only. 

The situation is, however, more involved for short chains. In case of short chains, the end
entropy contributes significantly in addition to the loop entropies. Hence the breaking of 
few bonds as well as the strand separation happens to be in a very narrow range of 
temperature. Thus average fraction $\theta(= 1 - \phi)$ of bonded base pairs 
is defined as \cite{wartell,campa,ns1},
\begin{equation}
\label{eq10}
\theta = \theta_{ext}\theta_{int} 
\end{equation}
$\theta_{ext}$ is the average fraction of strands forming duplexes, while $\theta_{int}$ 
is the average fraction of unbroken bonds in the duplexes. The equilibrium dissociation 
of the duplex $C_2$ to single strand $C_1$ may be represented by the relation 
$C_2 \rightleftharpoons 2C_1$. The dissociation equilibrium can be neglected in the case
of long chains; while $\theta_{int}$ and thus $\theta$ goes to zero while $\theta_{ext}$ 
is still practically 1. As discussed above, when $\theta$ goes practically from 1 to 0 at 
the melting, the two strands may not get completely separated, while for short 
chains, the single bond disruption and strand dissociation occur in a very narrow range
of temperature. Therefore, one need to compute both $\theta_{int}$ and $\theta_{ext}$.

To compute $\theta_{int}$, one has to separate the configurations describing 
a double strand on the one hand, and dissociated single strand on the other \cite{campa}. 
Since a bond between two bases is said to be broken if their separation is greater than 
the average separation between two bases. Therefore, $i^{th}$ bond is considered to be broken
if the value of $y_i$ is larger than a chosen threshold $y_0$. A configuration belongs 
to the double strands if at least one of the $y_i^{'s}$ is smaller than this $y_0$. 
One can therefore define $\theta_{int}$ for an $N$ base pair 
duplexes by:
\begin{equation}
\label{eq00}
\theta_{int} = \frac{1}{N} \sum_{n=1}^{N}\langle \vartheta(y_0 - y_i)\rangle
\end{equation}
where $\vartheta(y)$ is Heaviside step function and the canonical average 
$\langle.\rangle$ is defined considering only the double strand configurations. 
For $y_0$, we have taken a value of 2 ${\rm \AA}$. 

For $\theta_{ext}$ we use the expression given in refs. \cite{campa,ns1}.

\section{Temperature induced transition}
\label{temp}
When the dsDNA is in a thermal bath, due to thermal fluctuations the individual bonds
may disrupted. This cause the thermal melting of dsDNA. In this section we investigate
the role of salt concentration on the thermal stability of the dsDNA molecule. For most
of the thermal denaturation studies this effect has been ignored. Here we extend 
the previous studies on thermal denaturation of dsDNA using modified PBD model and
reproduce the experimental findings. We choose three chains, for which the experimental 
results are available \cite{owc}. These chains vary in terms of the fraction of GC \& AT 
base pairs. We call them as 30\% GC, 50\% GC and 75\% GC chains. The chains are, \\
(a) 5'-TGATTCTACCTATGTGATTT-3' (30\% GC) \\
(b) 5'-TACTTCCAGTGCTCAGCGTA-3' (50\% GC) \\
(c) 5'-GTGGTGGGCCGTGCGCTCTG-3' (75\% GC)\\
As the number of base pairs in all the three chains are 20, we consider them as short 
chains. We adjust the model parameters to match our results with the 
experiment \cite{owc}. It is found that the stiffness of the chain plays a crucial role in 
the melting or denaturation of the dsDNA molecule, in addition to the bond energy. While bond energy
is represented by Morse potential, the strand elasticityis represented by stiffness parameter $k$ 
in the anharmonic stacking term. Thus along with the value of $D$, we check various values of $k$ and $\rho$ 
to get close match the experimental results. We found the values of $\rho = 1.0$ 
and $\kappa = 0.01\; {\rm eV/\AA^2}$  as suitable choice for the current investigation.
\begin{figure}[t]
\begin{center}
\includegraphics[height=3.0in,width=4.0in]{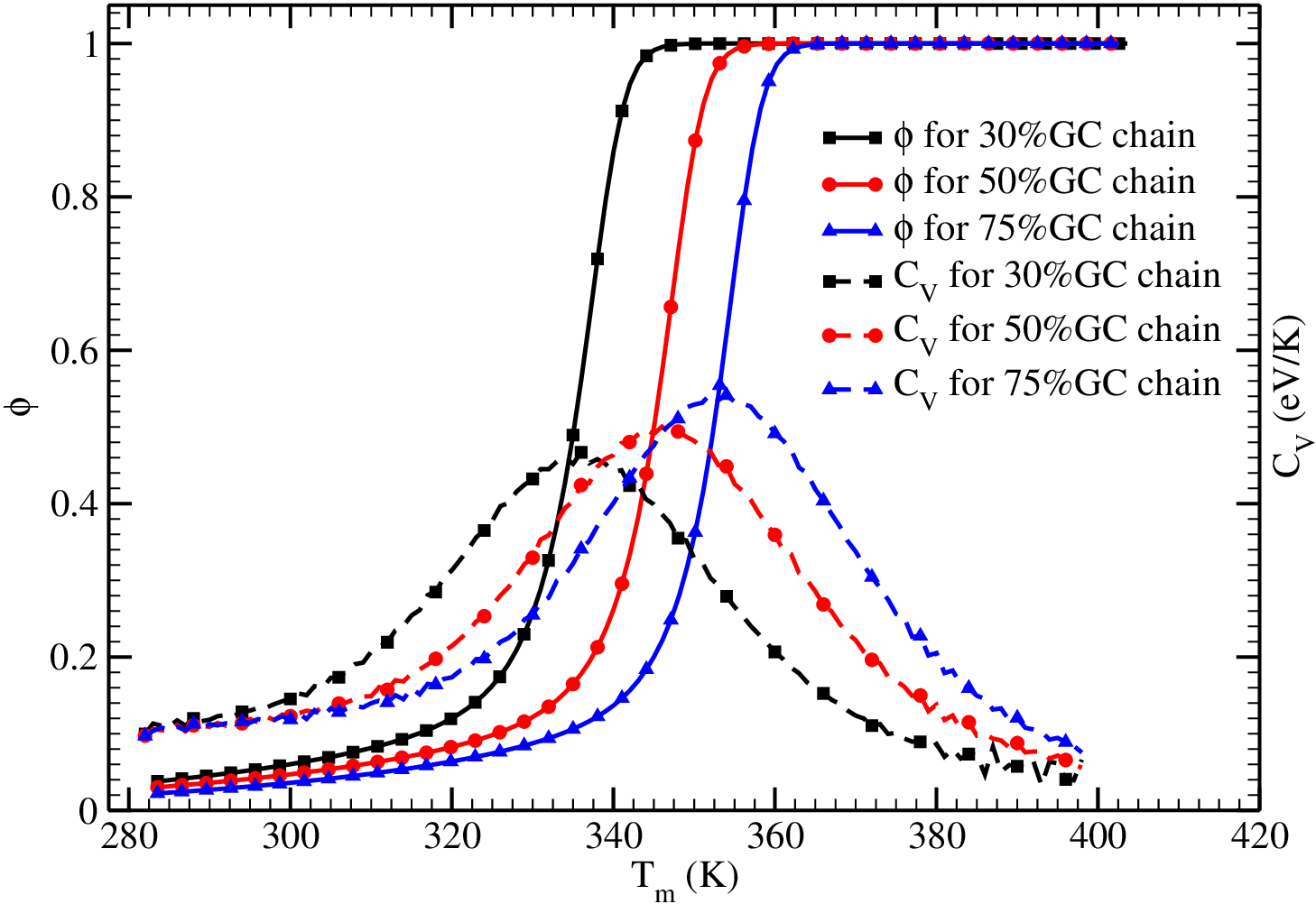}
\caption{The average number of open pairs as calculated using eq.(10) for salt concentration
of 0.621 M for all the three chains. As these are short chains, we calculate $\theta = \theta_{ext}
\theta_{int}$. The value of $C_v$ is scaled to show that the peak position and the 
50\% of the open pairs meet at the same point (temperature). }
\label{fig03} 
\end{center}
\end{figure}

To adjust the melting temperature within the range of the experimental observations, 
we finally tune the values of $D$, keeping other parameters same for all the three chains. 
With the value of $D$ (for AT base pair) as 0.090, 0.089 \& 0.087 eV for 30\% GC, 50\% GC 
\& 75\% GC chain respectively, we found close match with the experimental results for 
all the concentrations. The values of $D$ for GC base pair is 1.5 times of these values.
In the present investigation the stiffness parameters is considered as site independent, 
however, elasticity of the strand is suppose to depend on the distribution of different bases along 
the strand \cite{krueger,blake}. The effect of the stacking heterogeneities are left for further investigations.
The complete set of model parameters (except potential depth $D$) is: the inverse of potential depth, $a_{\rm AT} = 4.2 \; {\rm \AA^{-1}}$,
$a_{\rm GC} = 6.3 \;{\rm \AA^{-1}}$, single strand elasticity, $\kappa = 0.01\; {\rm eV/\AA^2}$,  
anharmonicity in the strand, $\rho = 1.0$,  range of anharmonicity, $b = 0.35 \;{\rm \AA^{-1}}$,
solution constant, $\lambda = 0.01$ and the solvent interaction factor $\gamma = 1.0 \; {\rm \AA^{-1}}$.

We calculate the free energy and the specific heat per base pair of the system 
using eq. (\ref{eqn8}) and eq. (\ref{eqn9}). At the temperature when the system gets the
sufficient amount of energy that is needed for transition from double stranded configuration
to single stranded configuration the free energy shows a kink. For better visualization, we
show the transition through the specific heat per base pair as a function of temperature.
At the transition point, this is shown by a peak. In order to avoid the overflow
we show the curve in fig. \ref{fig03} for only one value of concentration (0.621 M) for 
all the three chains. 
\begin{figure}[t]
\begin{center}
\includegraphics[height=3.0in,width=4.0in]{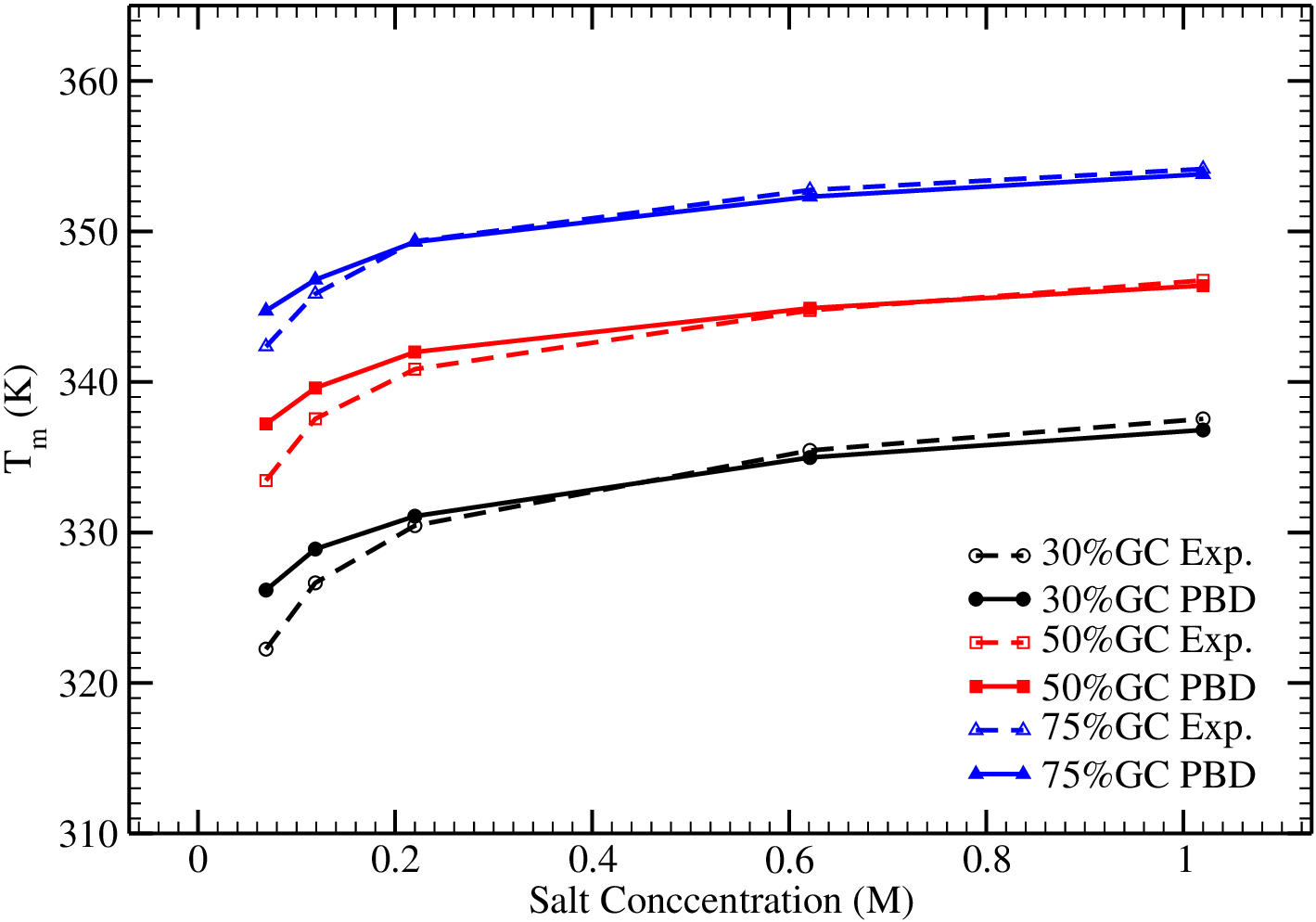}
\caption{Temperature-salt phase diagram showing the variation in $T_m$ as a 
function of salt concentration for all the three chains. The melting temperatures are 
evaluated at the concentrations, 0.069 M, 0.119 M, 0.220 M, 0.621 M \& 1.02 M as the 
experimental data for these values are available \cite{owc}.}
\label{fig04}
\end{center}
\end{figure}
We also monitor the fraction of open pairs as a function of temperature using 
eq. (\ref{eq10}). For short chains, the breaking of hydrogen bonds and the strand 
dissociation occurs in the same temperature range, thus, one has to calculate the
the $\theta_{ext}$ as well as $\theta_{int}$ \cite{campa}. The chain is said to be 
denatured when 50\% of the base pairs are in open state. The temperature corresponding 
to $\theta = 0.5$ is same as we get from the calculation of specific heat 
(fig. \ref{fig03}). Although the nature of all the three curves are different, they show 
the transition from double stranded configuration to single stranded configuration of DNA molecule. 
We obtain the value of 
melting temperature, $T_m$, for all the five concentrations for which the experimental 
results are available. The results are shown in fig. \ref{fig04}.
The phase diagram in fig. \ref{fig04}, shows the variation in the melting temperature 
as a function of salt concentration of the solution. It has been observed, experimentally as
well as theoretically, that the melting temperature has a logarithmic dependence
on the concentration of Na$^{+}$ in the solution. The value of potential depth, $D$, 
is tuned to get the proper match with the experimental data. The results shown are close to
the experiments, however at low value of salt concentration our data points have slight deviation
($\sim$ 2-3 K) from the experimental data. We observe that the deviation from the experimental
data at low concentration, is most for weaker chain (30\% GC). At the lower concentration 
the melting of AT \& GC pairs differs significantly \cite{blake} and the stiffness parameter 
may play an important role. Further investigation is required to explore the stacking energy 
as a function of the salt concentration in order to get closer match with the experimental 
results at the lower concentrations . 

\section{Force induced transition}
\label{force}
In this section we investigate the role of salt concentration on the mechanical unzipping 
of dsDNA molecule. In {\it vitro}, the double stranded DNA is pulled mechanically,
keeping other end fixed. These experiments are performed either at constant 
displacement of end pairs \cite{ns3,bock} and calculating the force required by the derivative of
the work done in the process or at varying loading rates \cite{danil}. 
Although both the set-ups give the same critical force for an infinite chain, the microscopic 
and dynamic behavior of unzipping of dsDNA in the the two ensembles are different. 
When the displacement is held constant, the force adjusts to compensate for the different 
average binding energies in AT-rich and GC-rich regions. This ensemble known as constant 
extension ensemble (CEE). The force required to break a pair fluctuate around the value of 
critical force, $F_c$. In this case, the large jumps and metastable states are usually absent. 
In the other ensemble that is constant force ensemble (CFE), the dsDNA is unzipped by 
applying constant force on one end of the strend keeping other end fixed. For homopolymeric DNA, the 
unzipping transition is smooth with the constant applied rate, once the constant applied 
force exceeds the threshold for separating the single base pairs. However, for heterogeneous 
chain, the transition from double stranded to single stranded is not smooth, but having 
several pauses and jumps depending on the distribution of weak (AT) and strong (GC) pairs.
\begin{figure}[t]
\begin{center}
\includegraphics[height=3.0in,width=4.0in]{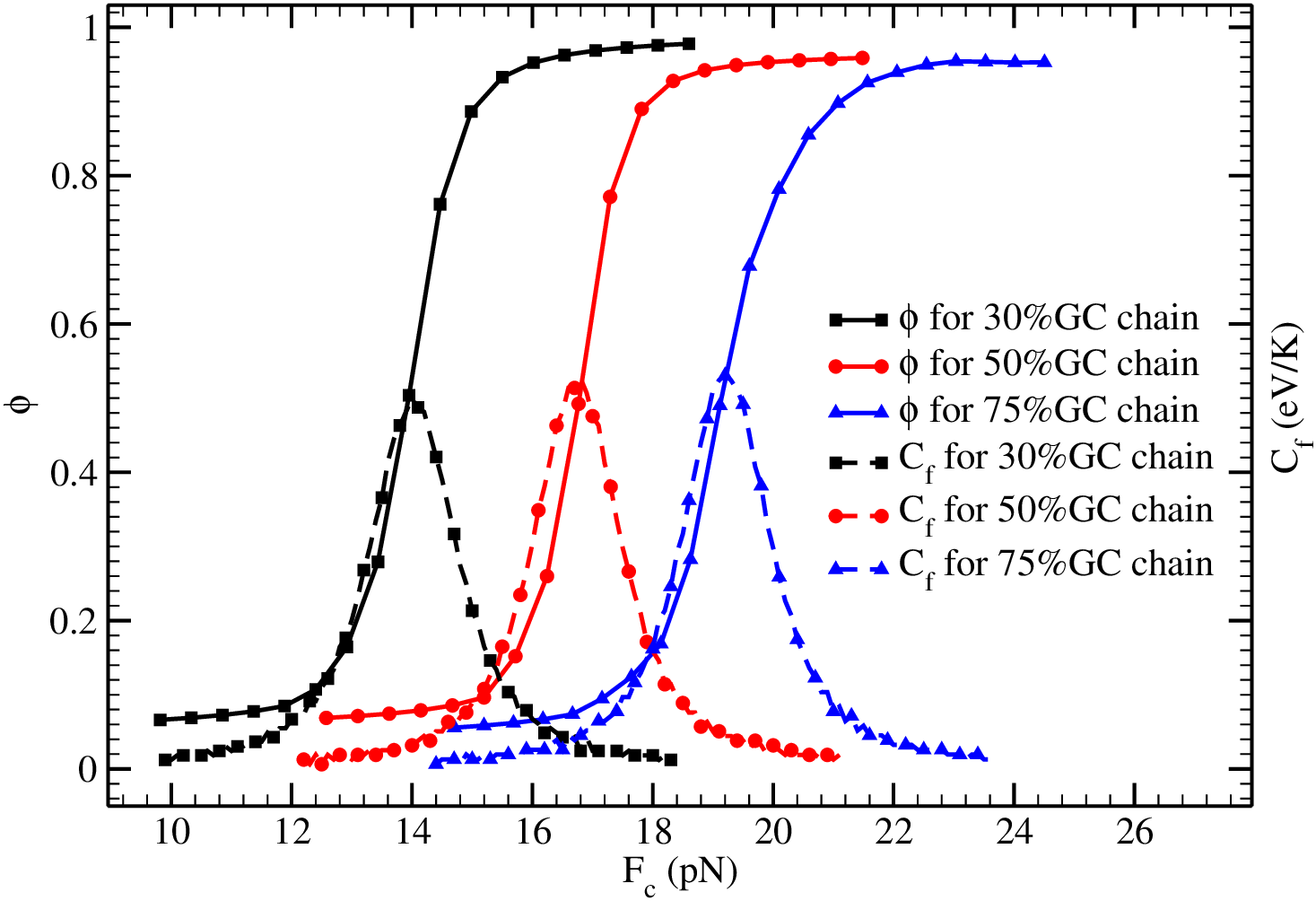}
\caption{The average number of open pairs as a function of the applied force 
on the chain for the salt concentration of 0.621 M. As this is for an infinite
chain, the $\theta \approx \theta_{int}$. Again value of specific heat is
scaled to get the two curves on the same plot.}
\label{fig05}
\end{center}
\end{figure}
In the current investigation we calculate the unzipping force in the constant force 
ensemble \cite{somen,nelson}. We take the same sequence of 20 base pairs which
we considered for thermal studies, however we repeat them to a length which can be 
considered as an infinite chain. We found that a length of about 600 base pairs is 
sufficient to be considered as an infinite chain. This length may be model dependent and 
one may get different number of base pairs in a chain that can be considered as infinite 
chain.  The modified Hamiltonian of the system is,
\begin{eqnarray}
\label{eqn1}
H_f = H - F\cdot y_e
\end{eqnarray}
We include a term $F\cdot y_e$ in eq. (\ref{eqn1}), as the force is applied on the end pair. 
The other model parameters are taken same as in the previous section, however, we tune
the value of potential depth $D$ in such a way that we get closer to the experimental results 
\cite{huguet} for 50\% GC chain. We take the value of potential depth as $D_{AT}=0.076 \; {\rm eV}$ 
for all the three chains. As the force induced unzipping experiments are performed at room temperature
($\sim$300 K), the melting temperature $T_m$ of the system should be much higher than 300 K, in order to ensure 
that at 300 K, base-pair opening is not due to thermal fluctuations. 
With these sets of model parameters $T_m$ is approximate 350 K for 50\% GC chain.
\begin{figure}[t]
\begin{center}
\includegraphics[height=3.0in,width=4.0in]{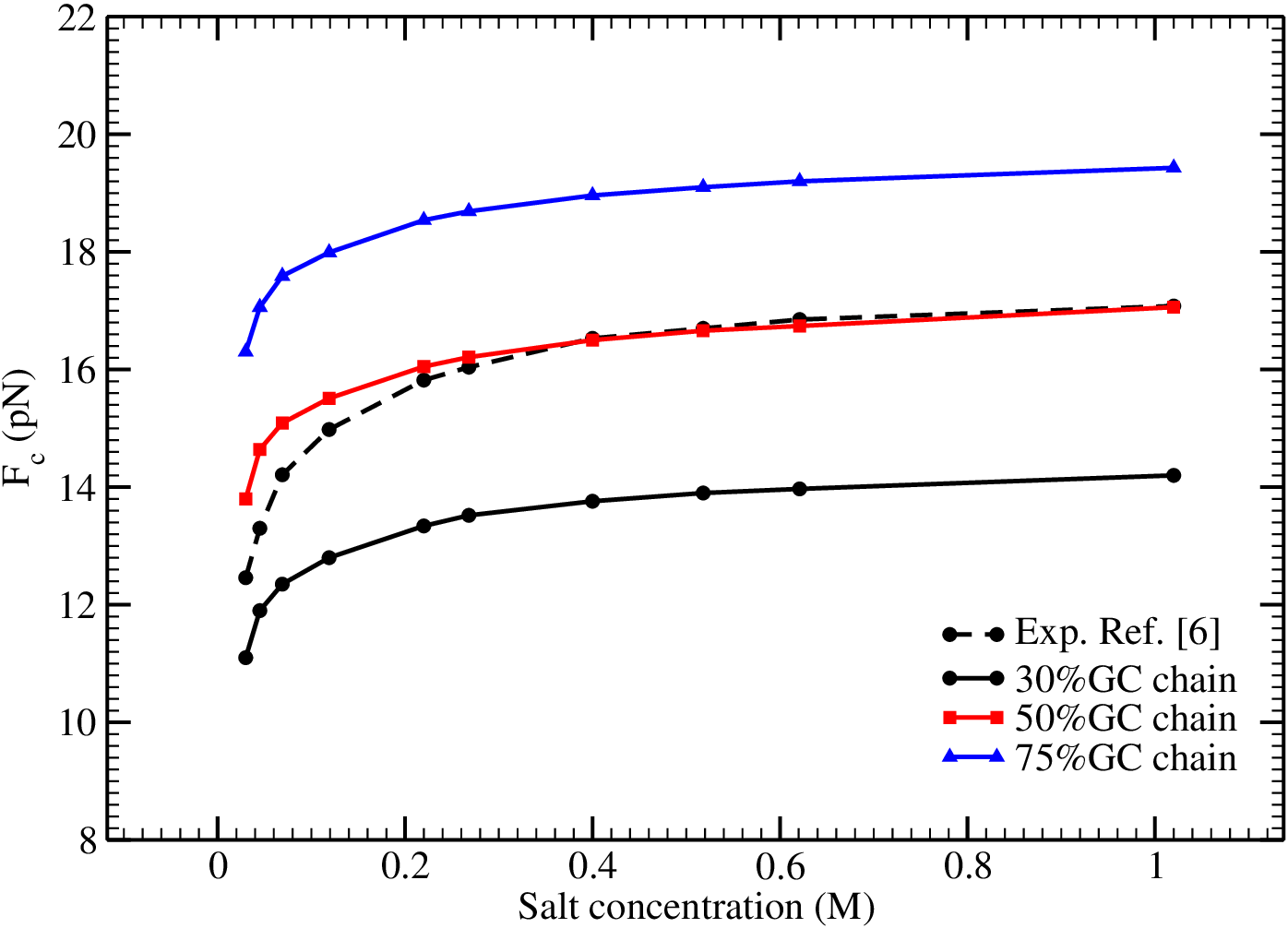}
\caption{The phase diagram showing the dependence of critical force on the salt concentration
of the solution. For comparison we show the experimental results obtained by \cite{huguet}, 
in the diagram.}
\label{fig06}
\end{center}
\end{figure}
We calculate the free energy of the system as a function of applied force. 
The force is applied on the 3' end of the chains at temperature 300 K. At the critical force 
we observe a kink in the free energy which gives rise to the peak in the specific heat at constant force
as shown in fig. \ref{fig05}. This is the point where the system transform from close state to open state. 
We obtain the value of critical force for all the three chains by locating the peak in specific heat 
as a function of applied force. We also calculate the average number of open pairs in a chain as a 
function of the applied force. From the fig. \ref{fig05}, it is clear that the $\phi = 0.5$ 
at the same value of force as predicted by specific heat as a function of applied force. As the 
calculation for force induced unzipping is for infinite chain, we calculate the 
$\phi = 1 - \theta = 1 - \theta_{int}$. 

The phase diagram in fig. \ref{fig06}, shows the two phases in case when the dsDNA is
forced to unzip mechanically as a function of salt concentration. We compare our results 
with the experimental phase diagram \cite{huguet}. The results
reported here are in good match with the experimental results except for low concentrations,
where the slight deviation has been observed ($\sim$ 0.5-1.4 pN). As discuss earlier the stacking interaction
contributes significantly at lower salt concentration during forced induced unzipping of DNA chain. 

\section{Conclusions}
\label{con}
In this manuscript, we have investigated the role of salt concentration on the thermal as well 
as on the mechanical unzipping behavior of heterogeneous dsDNA molecule. The PBD model 
is modified to incorporate the salt as well as the solvent effect of the system.
Our results indicate a close match with the earlier observations on thermal denaturation
of dsDNA, which shows that the melting temperature varies non-linearly or logarithmically with the 
salt concentration of the solution. As predicted by Manning's counterion condensation theory 
\cite{manning}, this is due to the layer of condensed counterions on the DNA surface that neutralizes 
the phosphate charges. This decreases the inter-strand electrostatic repulsion and the overall stability of 
DNA molecule increases and hence system need more thermal energy 
to break or denaturate. The melting temperature is found to vary with the salt concentration as 
well as with the GC content of the chain. We have investigated the role of salt present in the 
solution on the mechanical unzipping behavior of the dsDNA molecule in CFE.
We found that the critical force (force needed  to completely unzipped the molecule) 
increases with the increase in the salt concentration of the solution. The addition of salt 
in the solution basically shields the repulsion between the phosphate groups in the dsDNA 
chain which in result need more force to unzip the chain. Our results are found to be in
close agreement with the experimental phase diagram \cite{huguet}. However, the deviation at
the lower salt concentration needs further attention. The stacking heterogeneity, 
which has been taken as constant in the current investigation, might be the key factor at low concentration. 
We conclude that the PBD model, although
a quasi one dimensional model in nature, can be a good choice to investigate the presence of salt in the 
solution and its effect not only on the thermal denaturation of dsDNA chain, but also on the mechanical 
unzipping of the chain. 
In future, it would be interesting to study the 
dynamics of unzipping of dsDNA when the salt present in the solution is considered. 
However we would like to investigate this effect on the mechanical unzipping in the case
when force is not applied on the one of the ends but some where on the middle of the chain, the situation
 that is more closer to the transcription process.

\section*{Acknowledgement}
We are thankful to Yashwant Singh and Sanjay Kumar, Department of Physics, Banaras Hindu
University, India, for useful discussions and drawing our attentions to some experimental papers. We 
acknowledge the financial support provided by University Grant Commission, New Delhi, India.

\end{document}